\RequirePackage{fix-cm}

\documentclass[smallextended]{svjour3}       

\smartqed  

\usepackage{graphicx}

\usepackage{mathptmx}      

\usepackage{url}
\usepackage{float}
\begin{document}

\title{The Cryogenic AntiCoincidence detector for ATHENA X-IFU: a scientific assessment of the observational capabilities in the hard X-ray band.
\thanks{The final publication is available at Springer via http://dx.doi.org/10.1007/s10686-017-9543-4}
}

\titlerunning{The CryoAC for ATHENA X-IFU: observational capabilities in the hard X-ray band}       

\author{M. D'Andrea \and S. Lotti \and C. Macculi \and L. Piro \and A. Argan \and F. Gatti }
\institute{Matteo D'Andrea \at
              INAF/IAPS Roma, Via del Fosso del Cavaliere 100, 00133 Roma (Italy) \\
              Dept. of Physics, Univ. of Roma ''Tor Vergata'', Via della Ricerca Scientifica 1, 00133 Roma (Italy) \\
              Tel.: +390649934379\\
              \email{matteo.dandrea@iaps.inaf.it}      
           \and
           A. Argan \and S. Lotti \and C. Macculi \and L. Piro \at
              INAF/IAPS Roma, Via del Fosso del Cavaliere 100, 00133 Roma (Italy)
           \and
           F. Gatti \at
              Dept. of Physics, Univ. of Genova, Via Dodecaneso 33,16146 Genova (Italy)
}

\date{Received: date / Accepted: date} 

\maketitle

\begin{abstract}

ATHENA is a large X-ray observatory, planned to be launched by ESA in 2028 towards an L2 orbit. One of the two instruments of the payload is the X-IFU: a cryogenic spectrometer based on a large array of TES microcalorimeters, able to perform integral field spectrography in the 0.2-12 keV band (2.5 eV FWHM at 6 keV). The X-IFU sensitivity is highly degraded by the particle background expected in the L2 orbit, which is induced by primary protons of both galactic and solar origin, and mostly by secondary electrons. To reduce the particle background level and enable the mission science goals, the instrument incorporates a Cryogenic AntiCoincidence detector (CryoAC). It is a 4 pixel TES based detector, placed $<$1 mm below the main array. In this paper we report a scientific assessment of the CryoAC observational capabilities in the hard X-ray band (E$>$10 keV). The aim of the study has been to understand if the present detector design can be improved in order to enlarge the X-IFU scientific capability on an energy band wider than the TES array. This is beyond the CryoAC baseline, being this instrument aimed to operate as anticoincidence particle detector and not conceived to perform X-ray observations.

\end{abstract}

\section{Introduction}
\label{sec:1}
The Advanced Telescope for High Energy Astrophysics (ATHENA) \cite{athena} is the second large-class mission in the ESA Cosmic Vision 2015-2025, scheduled to be launched in 2028 towards an L2 orbit. The mission will address the science theme ``The Hot and Energetic Universe'' \cite{heuniverse}, studying some of the most pressing topics in contemporary Astrophysics through X-rays observations. The X-ray Integral Field Unit (X-IFU) \cite{XIFU} is one of the two instruments of the payload. It is a cryogenic spectrometer based on a large array of Transition Edge Sensor (TES) microcalorimeters, able to perform simultaneous high-resolution energy spectroscopy in the 0.2-12 keV band ($\Delta$E = 2.5 eV at 6 keV) and imaging over a 5 arcmin diameter FoV (with 5'' pixels). The X-IFU sensitivity is highly degraded by the particle background expected in L2 orbit, which is induced by primary protons of both solar and Cosmic Rays origin, and by secondary electrons. To meet the instrument scientific goals, enabling the characterization of faint or diffuse sources (e.g. WHIM or Galaxy cluster outskirts), it is necessary to reduce this background by a factor $\sim$100 down to the requirement of  0.005 cts/cm$^2$/s/keV. This can be achieved combining a passive electron shielding with the Cryogenic AntiCoincidence detector (CryoAC) \cite{CryoAC}, which is a 4 pixels TES-based detector placed within a proper optimized environment surrounding the X-IFU TES array.

In this paper we will report a feasibility study performed to explore the observational capabilities of the CryoAC detector in the hard X-ray band (E $>$ 10 keV). The aim of the study is to understand if the present detector design can be improved in order to enlarge the scientific capability of the X-IFU instrument over a wider energy bandwidth. We remark that this is beyond the CryoAC baseline, being this instrument not aimed to perform X-ray spectroscopy but conceived to operate as anticoincidence particle detector. We will start with a brief overview of the CryoAC detector, in order to show the baseline configuration that represents the starting point for this study (Sect.~\ref{sec:2}). Then, we will evaluate the detector limit fluxes (Sect.~\ref{sec:3}) and will examine a case study of its scientific performance taking into account astronomical sources (Sect.~\ref{sec:4}). The outcome of this scientific assessment will be the detector requirements enabling the CryoAC high energy observational capabilities, which will be reported in the conclusions (Sect.~\ref{sec:5}).

\section{The Cryogenic AntiCoincidence detector: an overview}
\label{sec:2}

\begin{figure}[H]\sidecaption
\resizebox{0.5\hsize}{!}{\includegraphics{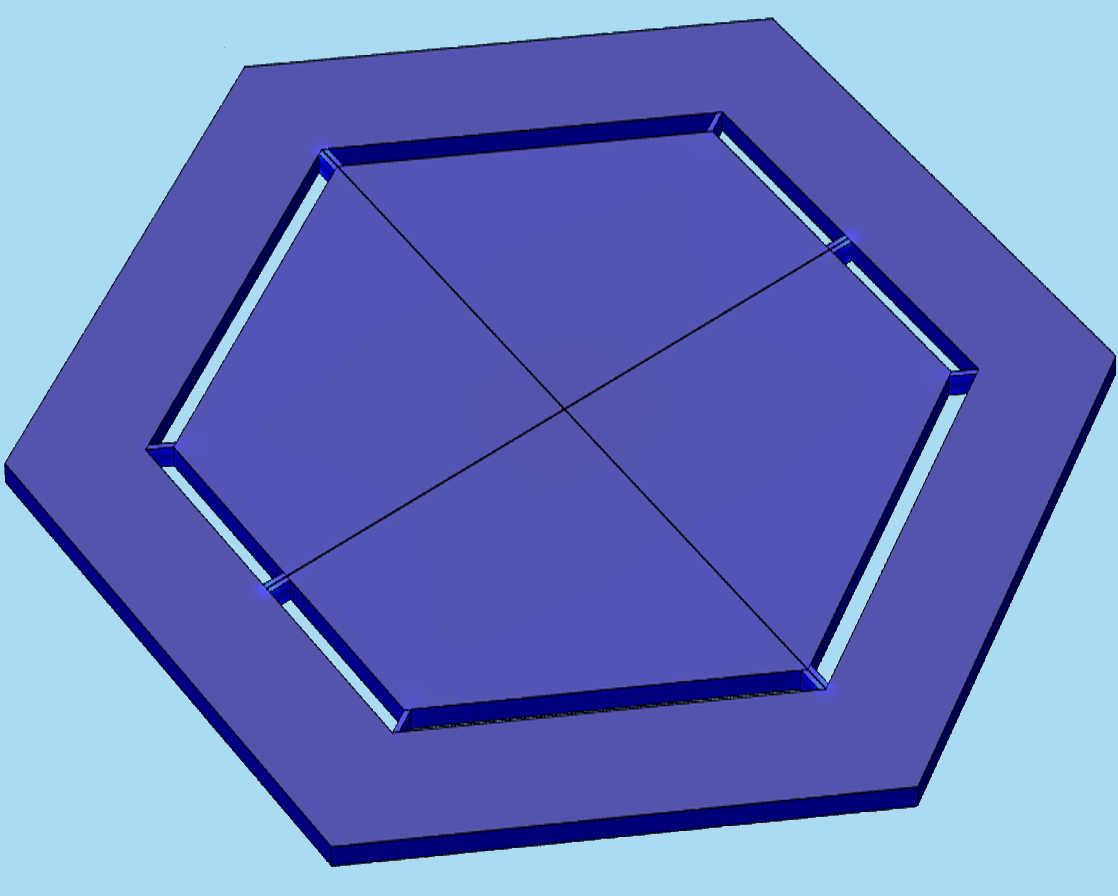}}
\caption{The CryoAC detector schematic: 4 trapezoidal silicon pixels connected to a silicon rim by 3 bridges per pixel. On each absorber will be deposited a network of about 100 TESes (here not shown).}
\label{fig:cryoac}
\end{figure}

The CryoAC detector is placed $<$1 mm below the X-IFU TES array. The baseline design is based on absorbers made of a single Silicon layer (500 $\mu$m thick), where the energy deposited by particles is sensed by Iridium TES sensors. The detector has a bandpass of 5 keV - 750 keV (TBC) and its active part covers a full area of 4.6 cm$^2$, larger than the main array (2.6 cm$^2$). For redundancy/deadtime issues the detector is divided into 4 indipendent trapezoidal pixels, each one having an area of 1.15 cm$^2$ and a separated readout chain. These pixels are inserted in an hexagonal silicon frame (i.e. the ''rim'') mechanically anchored to the cold stage (50 mK) of the Focal Plane Assembly. Each pixel is connected to the rim by means of 3 Silicon bridges, in order to realize a well defined and reproducible thermal conductance towards the thermal bath (see Fig.~\ref{fig:cryoac}). Each readout chain is based on a SQuID device operating in a standard FLL configuration. See \cite{XIFU}, \cite{CryoAC} and refs. therein for a more detailed scenario.

\section{Evaluation of the CryoAC instrumental sensitivity}
\label{sec:3}

To assess the observational capabilities of an X-ray detector the fundamental quantity to evaluate is its \textit{minimum detectable flux}. This parameter characterizes the detector performances in terms of signal to noise ratio, determining the instrumental sensitivity and then the suitable astronomical targets. We recall that the minimum detectable flux $F_{min}$ (ph/cm$^2$/s/keV) within an observing time $t$ (s) and a $\Delta E$ energy band (keV) can be expressed as \cite{fraser}:
\begin{equation}
F_{min} = \frac{n_\sigma}{A_{eff} Q} \sqrt{\frac{B_p A_d + Q j_d \Omega A_{eff}}{t \Delta E}}
\label{eqn:fmin1}
\end{equation}
where $n_\sigma$ is the desired confidence level, $A_{eff}$ is the X-ray optics effective area (cm$^2$), Q is the detector response function (counts/ photons),  $A_d$ its geometric area (cm$^2$), $B_P$ is the internal particle background level (cts/cm$^2$/s/keV), $j_d$ is the flux of the diffuse component of the background (ph/cm$^2$/s/keV/sr) and $\Omega$ is the detector aperture (sr). In the next sub-sections we will evaluate all the parameters reported above, considering the baseline CryoAC design.

\subsection{ATHENA mirror effective area}
\label{sec:3-2}

\begin{figure}[H]
\centering
\includegraphics[width=0.6\textwidth]{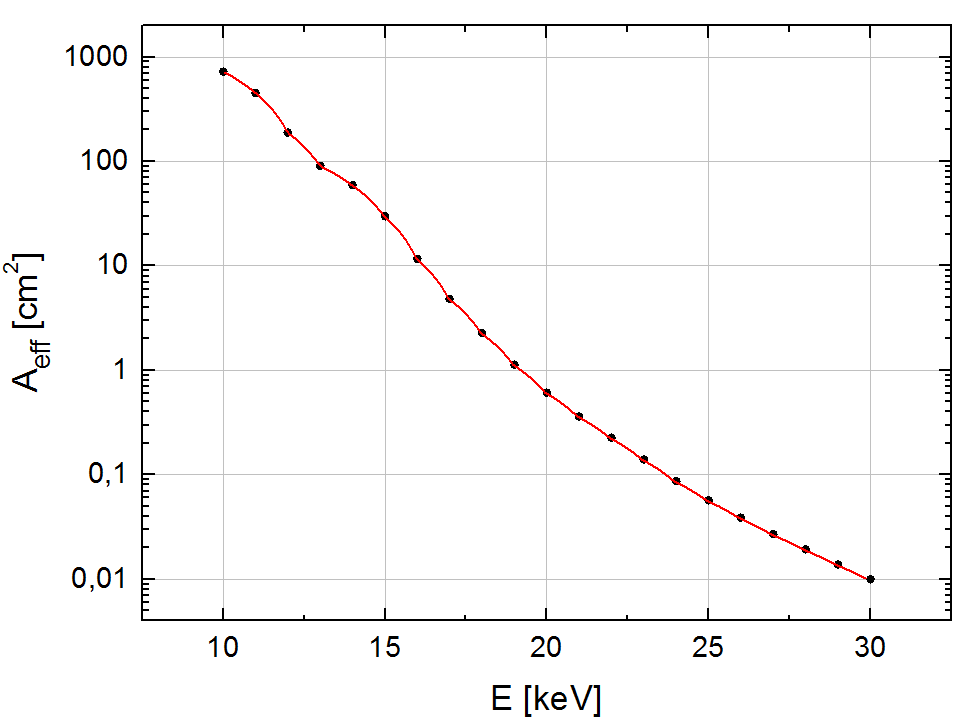}
\caption{On-axis effective area of the ATHENA telescope. The red line is an interpolation of the data courtesy of R. Willingale (black points).}
\label{fig:aeff} 
\end{figure}

The on-axis effective area of the ATHENA mirror in the range 10-30 keV is shown Fig.~\ref{fig:aeff}. The data has been provided by R. Willingale (private communication) and refers to optics with a mirror module radius R$_{max}$ = 1469 mm, 2.3 mm of rib spacing and a surface roughness of 5 \AA. For information about the ATHENA optics and their development see \cite{optics}.

Fig.~\ref{fig:aeff} shows that above 10 keV the mirror area rapidly drops, hitting the value of 1 cm$^2$ around an energy of 19 keV. Based upon this data, we can already assess that within the ATHENA context the target hard X-ray band is limited in the range from 10 keV to $\sim$ 20 keV. 

\subsection{CryoAC response function}
\label{sec:3-1}

We define the CryoAC response function as the probability that an X-ray photon focused by the ATHENA optics will be detected by the CryoAC. This function includes several contributions as the trasmissivity of the X-IFU thermal filters, the trasmissivity of the TES array and the quantum efficiency of the CryoAC (i.e. the photoelectric absorption efficiency in the absorber).

To evaluate the CryoAC response we build a Monte Carlo simulation using the Geant4 toolkit \cite{geant4}. The simulation is based on a simplified model of the X-IFU Focal Plane Assembly (FPA), which includes the five thermal filters (with a total thickness of 280 nm Polyimide and 210 nm Al), the TES array consisting of 3840 pixels of 249 $\mu$m pitch (absorber composition is 1 $\mu$m of Au covered by 4 $\mu$m of Bi) and the CryoAC detector in the baseline design (500 $\mu$m thick Silicon absorber).

\begin{figure}[H]
\centering
\includegraphics[width=0.49\textwidth]{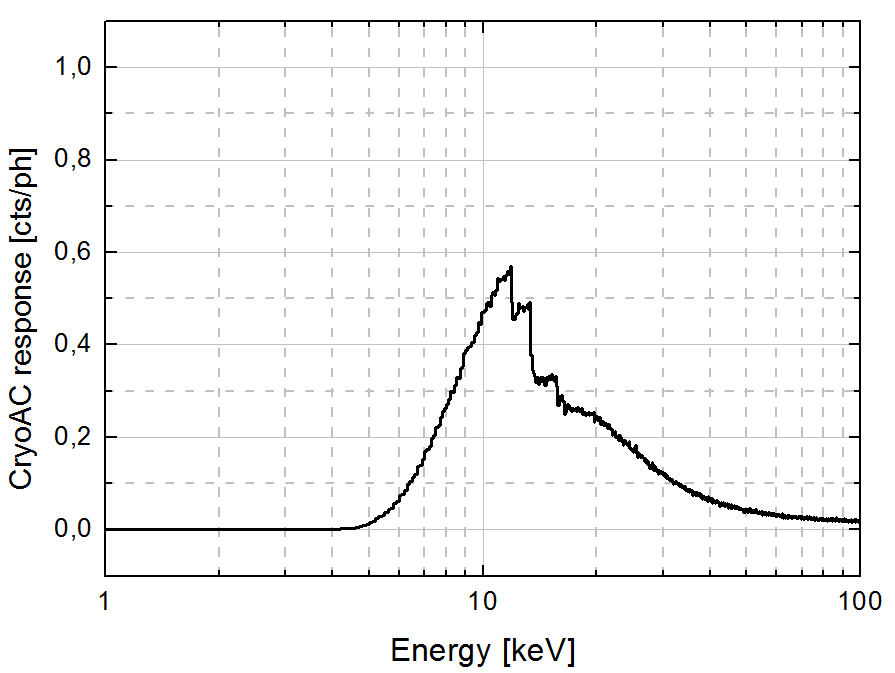}
\includegraphics[width=0.49\textwidth]{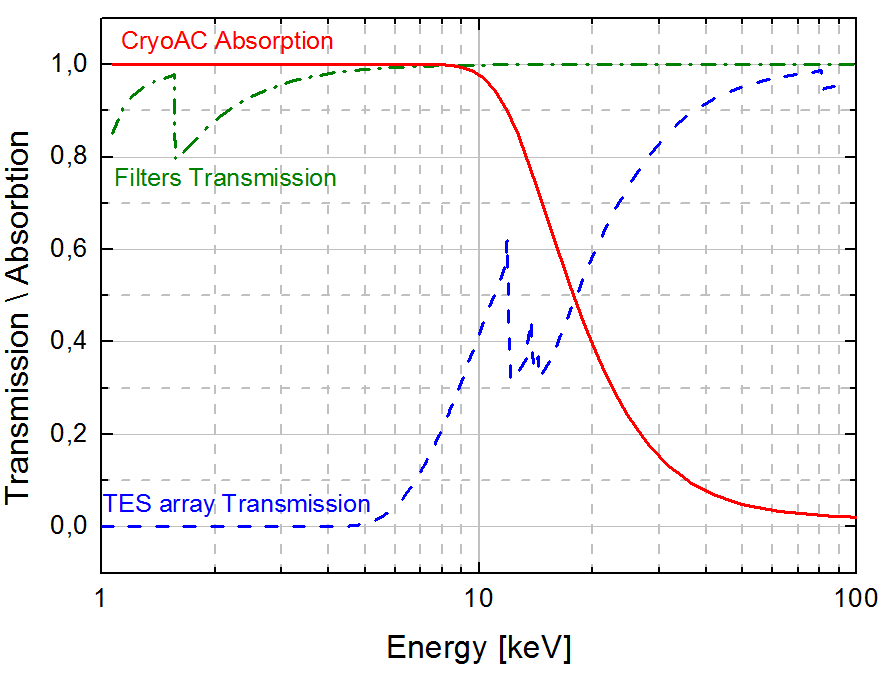}
\caption{(\textit{Left}): CryoAC response function estimated by means of a Geant4 simulation. (\textit{Right}): Different contributions to the CryoAC response function estimated using the attenuation coefficients tabulated by NIST \cite{nist}.}
\label{fig:Q} 
\end{figure}

The result of the simulation is shown in Fig.~\ref{fig:Q} \textit{left}. The response function rapidly grows at low energy, showing a maximum around 12 keV (Q$_{max}$ = 0.57 cts/ph), and decreases at higher energies, dropping below the Q = 0.10 cts/ph level above 30 keV. To understand the contribution of the different elements in the FPA to the response function, in Fig.~\ref{fig:Q} \textit{right} are reported the respective transmission/absorbtion curves estimated using the X-ray attenuation coefficients tabulated by NIST \cite{nist}.

Note that in the reference energy range from 10-20 keV (see Sect.~\ref{sec:3-2}) the current CryoAC design ensure a good absorbtion efficiency ($\sim$ 40\% at 20 keV), and so we do not consider useful to put effort into an upgrade of the absorber characteristics (i.e. the thickness or the material) at this stage.

\subsection{In orbit background}
\label{sec:3-3}

The diffuse component of the background has been estimated starting from the hard X-ray spectrum of the Cosmic X-ray Background (CXB) measured by IBIS INTEGRAL. It has been modeled using the analytical description proposed by Turler et al. in \cite{xback}. To obtain the background level we have folded this spectral flux with the optics effective area (Sect.~\ref{sec:3-1}), the detector response function (Sect.~\ref{sec:3-2}) and the aperture of a single CyoAC pixel $\Omega = 0.80 \cdot 10^{-6}$ sr (corresponding to a FoV of 9.4 arcmin$^2$).

The internal particle background has been instead estimated by means of the Monte Carlo simulations already developed to study the X-IFU in-orbit background and drive the design of the CryoAC. These simulations reproduce the L2 environment and the ATHENA mass model, tracking both the primary and secondary particles components (see \cite{lotti} and refs. therein for a detailed discussion). In this case we have taken into account the use of the X-IFU TES array as ``reverse'' anticoincidence device. This is the opposite of what happens in observations with the X-IFU array, where it is the CryoAC that discriminates main detector events that happen in both detectors. In this context the main detector can act as an anticoincidence device for the CryoAC, reducing the particle background to some extent despite not being designed/optimized for the scope. The residual particle background level on a CryoAC pixel is roughly constant in the 10-30 keV energy band, counting about 0.012 cts/s/keV.

\begin{figure}[H]
\centering
\includegraphics[width=0.6\textwidth]{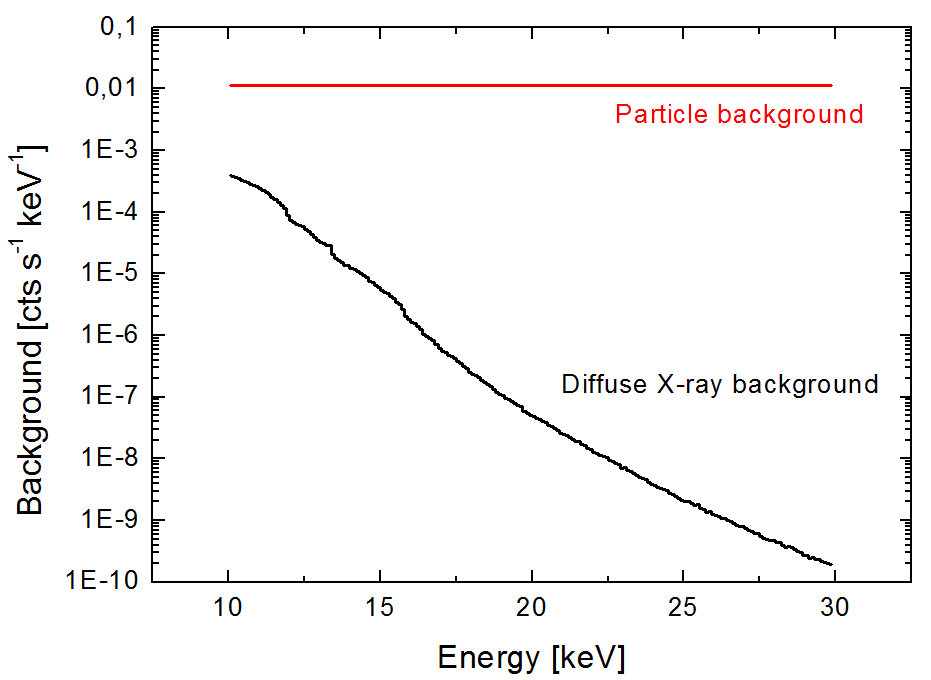}
\caption{Spectra of the particle and the diffuse X-rays backgrounds expected on a single CryoAC pixel in L2 orbit.}
\label{fig:background} 
\end{figure}

The estimated background spectra for a single CryoAC pixel in the band 10-30 keV are shown in Fig.~\ref{fig:background}. Note that above 10 keV the diffuse X-Ray component of the background is always negligible with respect to the particles one, and so it can be neglected.

\subsection{Limit fluxes and continuum sensitivity}
\label{sec:3-4}

\begin{figure}[H]
\centering
\includegraphics[width=0.49\textwidth]{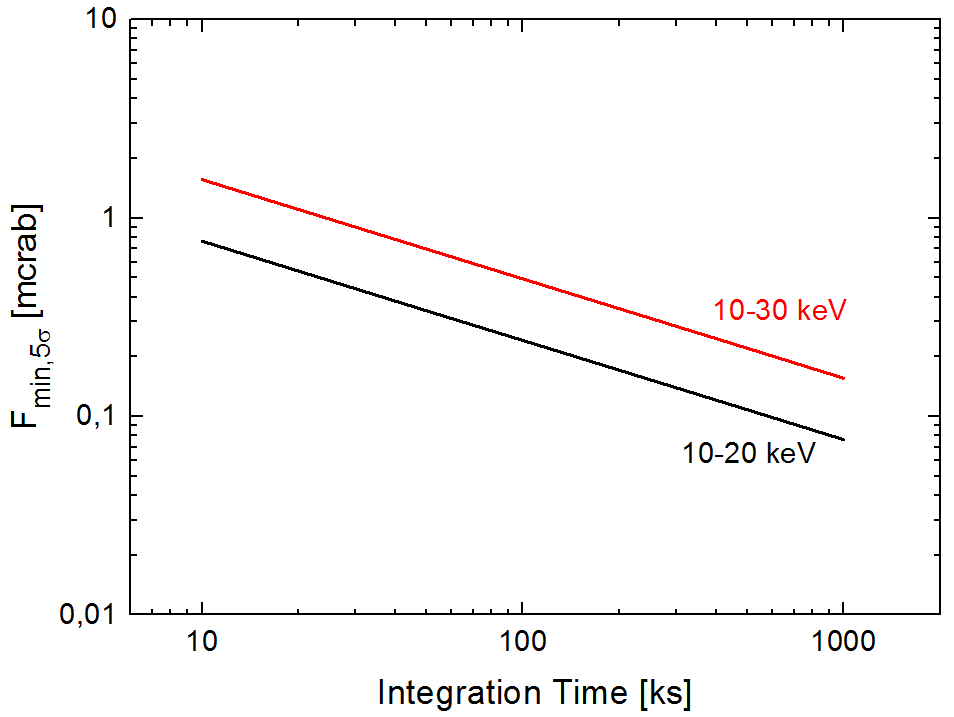}
\includegraphics[width=0.49\textwidth]{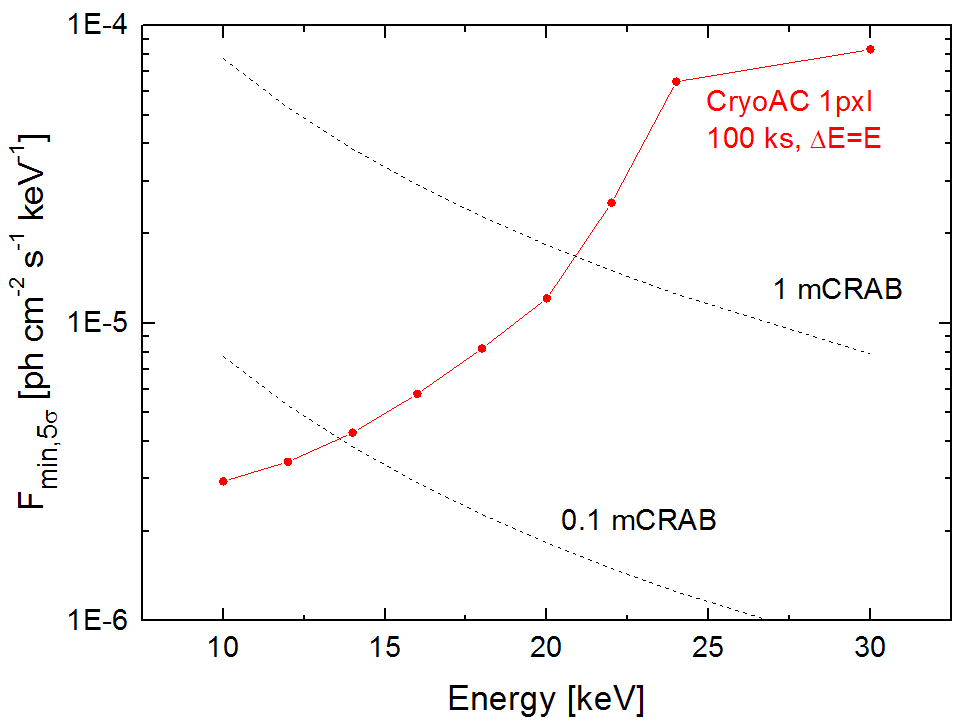}
\caption{(\textit{Left}): CryoAC minimum detectable flux as function of the observation time. (\textit{Right}):CryoAC continuum sensitivity for 100 ks of exposure time and an energy bandwidth $\Delta$E=E. Crab fluxes are overplotted as reference.}
\label{fig:fmin} 
\end{figure}

Including in eq. (1) the energy-dependence of the effective area and the response function, we have then estimated the minimum detectable flux for a single CryoAC pixel and a 5$\sigma$ confidence level (n$_\sigma$ = 5). In Fig.~\ref{fig:fmin} \textit{left} the limit flux is reported as a function of the observation time in two different energy ranges, whereas in Fig.~\ref{fig:fmin} \textit{right} is shown as a function of energy, considering in eq. (1) an energy range $\Delta$E=E and a 100 ks exposure for each point. These two plots characterize the sensitivity of the CryoAC pixels to a continuum emission, showing that the limit fluxes with the current detector baseline are fractions of mCrab in the band 10-20 keV. 

To better evaluate the sensitivity of the detector, in Tab.~\ref{tab:fluxes} are reported the limit fluxes of the CryoAC compared with the reference ones for NUSTAR and ATHENA X-IFU (assuming t = 100 ks and n$_\sigma$ = 5).

\begin{table}[H]
\centering
\caption{Limit fluxes for the CryoAC compared with the reference ones for NUSTAR and ATHENA X-IFU (t = 100 ks, n$_\sigma$ = 5)}
\label{tab:fluxes}
\begin{tabular}{lllll}
\hline\noalign{\smallskip}
Instrument & Energy range & $F_{min}$ & $F_{min}$ & Notes and refs. \\
 & [keV] & [erg/cm$^2$/s] & [mCrab] &  \\
\noalign{\smallskip}\hline\noalign{\smallskip}
CryoAC & 10-20 & 1.6$\cdot 10^{-12}$ & 0.2 & 1 pixel \\
CryoAC & 10-30 & 6.3$\cdot 10^{-12}$ & 0.5 & 1 pixel \\
NUSTAR & 10-30 & 5.0$\cdot 10^{-14}$ & 0.4$\cdot 10^{-2}$ & \cite{nustar}\\
ATHENA X-IFU & 2-10 & 3.2$\cdot 10^{-16}$ & 1.6$\cdot 10^{-5}$ & Point source \cite{lotti2} \\ 
\noalign{\smallskip}\hline
\end{tabular}
\end{table}

The result of the first part of this scientific assessment is that the CryoAC can operate as hard X-ray detector without changes in the baseline design, supplying a moderate sensitivity in the 10-20 keV  band (Tab.~\ref{tab:fluxes}). We remark that this narrow energy band is limited by the drop of the optics effective area at high energies, and not by the detector features.

\section{Scientific Simulations}
\label{sec:4}

Once assessed the continuum sensitivity of the CryoAC, in this section we will present some scientific simulations of astronomical targets that we have performed to understand the spectroscopic capabilities of the detector. This will allow us to define an optimal reference value for the CryoAC required energy resolution. We point out that at present no requirement has been set for this parameter. The last CryoAC single pixel prototype has been tested with 60 keV photons, showing an energy resolution $\Delta$E = 11.3 keV (FWHM) \cite{ACS7} due to the test setup. The intrinsic resolution of the detector is estimated to be more an order of magnitude lower.

\subsection{Crab observation}
\label{sec:4-1}

\begin{figure}[H]
\centering
\includegraphics[width=0.7\textwidth]{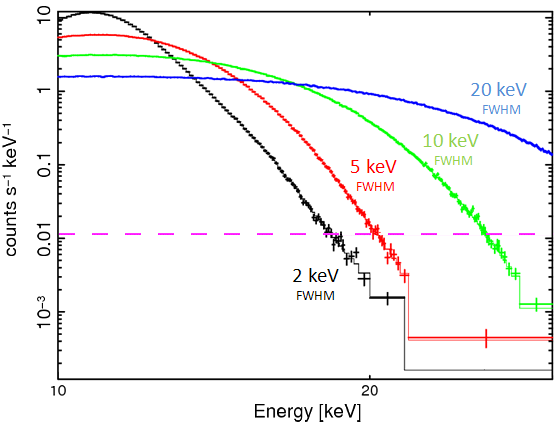}
\caption{Crab spectra for a 100ks observation with a CryoAC pixel and different values of the energy resolution. The violet dashed line represents the background level.}
\label{fig:crab} 
\end{figure}

To have a first reference, we have simulated with XSPEC \cite{xspec} a 100 ks observation of the Crab Nebula. The Crab spectrum has been modeled as a power law with photon index $\alpha$ = 2.08 and normalization K = 9.3 ph/cm$^2$/s at 1 keV \cite{crab}. Starting from the quantities evaluated in the previous sections, we have generated several CryoAC response matrices, varying the detector energy resolution from 2 to 20 keV (FWHM). Note that the energy resolution has been assumed constant in the range 10-30 keV. The Crab spectra simulated accounting for the different response matrices are shown in Fig.~\ref{fig:crab}. 

We have fitted the simulated dataset in order to assess the accuracy by which the source parameter could be recovered. 
The results of the fit are shown in Table \ref{tab:crab}, where we have reported the 90\% confidence range of the spectral parameter obtained with the different energy resolutions. As expected the spectral resolution deeply influences the observation, significantly conditioning the errors on the recovered parameters. Based upon these results, we can preliminary conclude that an energy resolution of few keV (FWHM) in the 10-20 keV range is necessary to extend the scientific capability of the CryoAC.

\begin{table}[H]
\centering
\caption{Spectral parameter of the Crab Nebula (90\% confidence range) obtained simulating a 100 ks observation with the CryoAC for different values of the energy resolution.}
\label{tab:crab}
\begin{tabular}{ccccc}
\hline\noalign{\smallskip}
Energy resolution& $\alpha$ & $\Delta \alpha /\alpha$ & K & $\Delta K / K$ \\
(FWHM) &  &  & [ph/cm$^2$/s/keV] &  \\
\noalign{\smallskip}\hline\noalign{\smallskip}
2 keV & 2.07 - 2.10 & 1.4\% & 9.1 - 9.8 & 7.2\%\\
5 keV & 2.06 - 2.12 & 2.8\% & 8.8 - 10.2 & 15\%\\
10 keV & 2.05 - 2.18 & 6.8\% & 8.6 - 12.0 & 36\%\\
20 keV & 2.10 - 2.51 & 20\% & 9.8 - 26.0 & 174\%\\
\noalign{\smallskip}\hline
\end{tabular}
\end{table}

\subsection{A case study: HMXBs observation}
\label{sec:4-2}

We want now to evaluate the CryoAC scientific capabilities and the role of its energy resolution in combination with the X-IFU TES array. Given the sensitivity estimated in Sect.~\ref{sec:3-4}, a target source with a flux of some tens of mCrab observed with an exposure time around 100 ks is needed to have enough counts on the CryoAC for spectral analysis. Bright High-Mass X-ray Binaries (HMXBs) can be considered an ideal candidate in this context. These sources can indeed present an high energy cut-off in the CryoAC energy band (10-20 keV), representing an interesting target for joint TES array/CryoAC observations.

\begin{figure}[H]
\centering
\includegraphics[width=0.7\textwidth]{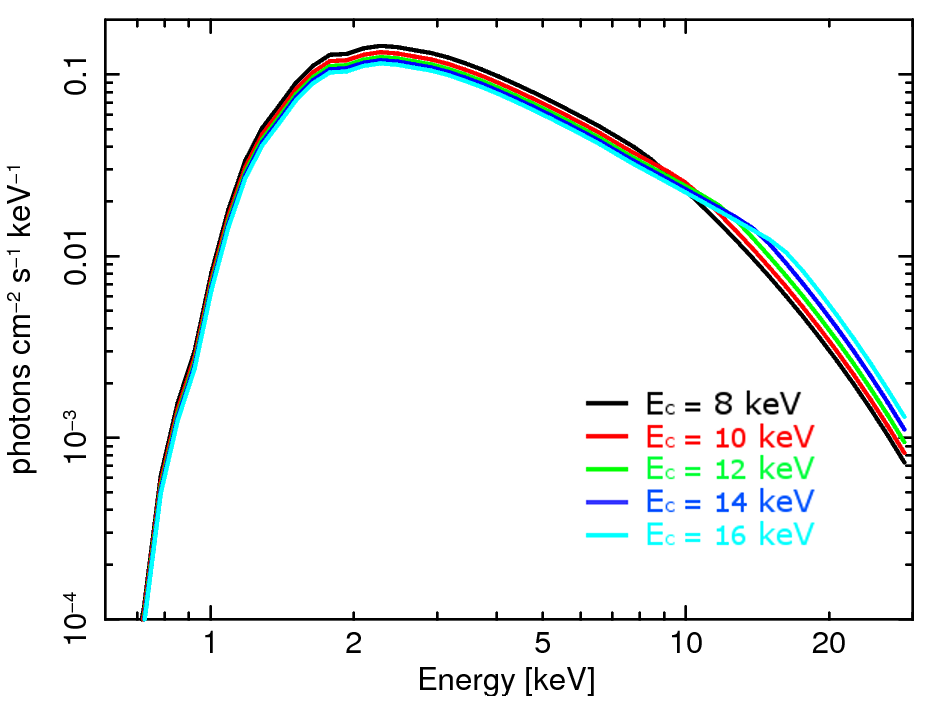}
\caption{The HMXB spectral models used in the simulations.}
\label{fig:hmxb} 
\end{figure}

As case study, we have therefore simulated the observation of High-Mass X-ray Binary spectra, exploring if the CryoAC could effectively extend the X-IFU energy band. For the HMXBs spectra we have used a simplified model with an absorbed power law spectrum and a high energy cut-off (\textit{wabs*highecut*powerlaw} in XSPEC), generating several models with different cutoff energies in the range 8-16 keV. The models are shown in Fig.~\ref{fig:hmxb} and their parameters are given in Table~\ref{tab:hmxbxtab}.

\begin{table}[H]
\centering
\caption{HMXB models parameters.}
\label{tab:hmxbxtab}  
\begin{tabular}{lll}
\hline\noalign{\smallskip}
Component & Parameter & Value \\
\noalign{\smallskip}\hline\noalign{\smallskip}
wabs & N$_H$ & 2$\cdot$10$^{22}$ cm$^2$\\
highecut & E$_C$ & 8, 10, 12, 14, 16 keV\\
highecut & E$_F$ & 10 keV\\
powerlaw & $\Gamma$ & 1.5\\
 & Flux 0.2 - 20 keV & 100 mCrab\\
\noalign{\smallskip}\hline
\end{tabular}
\end{table}

\begin{figure}[H]
\centering
\includegraphics[width=0.7\textwidth]{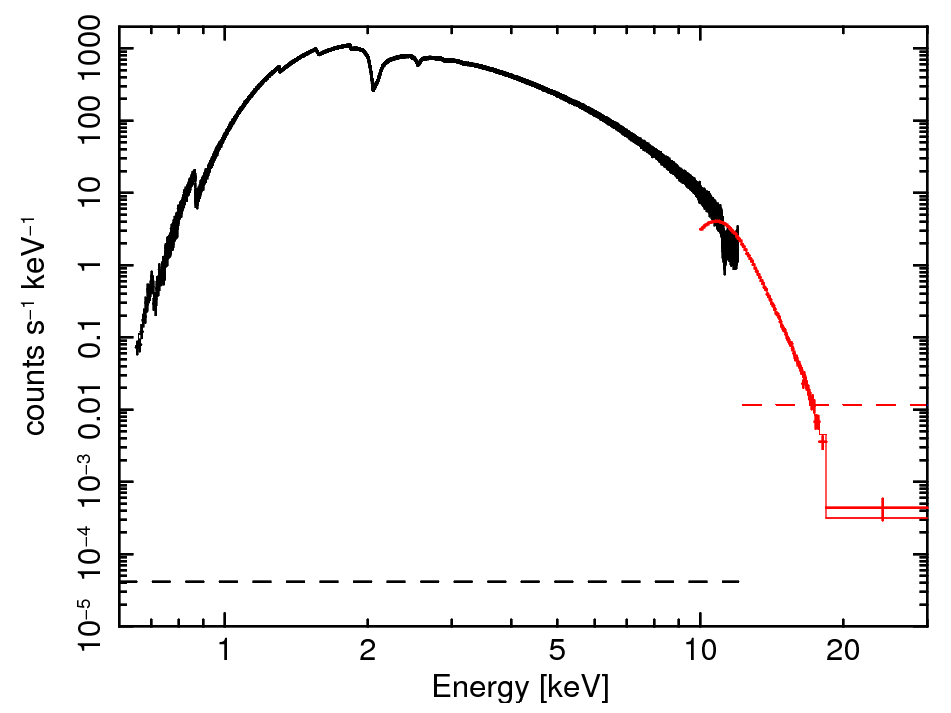}
\caption{HMXB spectrum with a cut-off at E$_C$ = 12 keV observed with both the X-IFU TES array (black data) and the CryoAC (red data) in 50 ks exposure time. Dashed lines represent the respectives background levels.}
\label{fig:hmxb2} 
\end{figure}

\noindent For each model we have simulated three different kind of observation: 
\begin{itemize}
\item 50 ks exposure with the X-IFU TES Array only
\item 50 ks exposure with both the X-IFU TES Array and the CryoAC (energy resolution $\Delta$E = 2keV FWHM)
\item 50 ks exposure with both the X-IFU TES Array and the CryoAC (energy resolution $\Delta$E = 5keV FWHM)
\end{itemize}
An example of a simulated spectrum is shown in Fig.~\ref{fig:hmxb2}.

\begin{table}[H]
\centering
\caption{High energy cut-off parameters obtained fitting the simulated spectra in the different configurations. The errors refer to the 90\% confidence level.}
\label{tab:hmxbtab2}      
\begin{tabular}{llll}
\hline\noalign{\smallskip}
Model & TES array &  TES array &  TES array \\
 &  & + CryoAC &  + CryoAC \\
& & ($\Delta$E = 2 keV) & ($\Delta$E = 5 keV) \\
\noalign{\smallskip}\hline\noalign{\smallskip}

&  &  &  \\
E$_C$ = 8 keV & E$_C$ = $8.01^{+0.02}_{-0.02}$ keV & E$_C$ = $8.00^{+0.02}_{-0.01}$ keV & E$_C$ = $8.01^{+0.02}_{-0.01}$ keV \\
& E$_F$ = $9.60^{+0.11}_{-0.11}$ keV & E$_F$ = $9.63^{+0.11}_{-0.11}$ keV & E$_F$ = $9.61^{+0.11}_{-0.11}$ keV \\
&  &  &  \\

&  &  &  \\
E$_C$ = 10 keV & E$_C$ = $9.97^{+0.04}_{-0.03}$ keV & E$_C$ = $9.93^{+0.03}_{-0.03}$ keV & E$_C$ = $9.94^{+0.03}_{-0.03}$ keV \\
& E$_F$ = $8.88^{+0.17}_{-0.43}$ keV & E$_F$ = $9.47^{+0.25}_{-0.24}$ keV & E$_F$ = $9.31^{+0.35}_{-0.35}$ keV \\
&  &  &  \\

&  &  &  \\
E$_C$ = 12 keV & E$_C$ = $9.2^{+0.3}_{-0.5}$ keV & E$_C$ = $11.6^{+0.2}_{-0.2}$ keV & E$_C$ = $11.0^{+0.2}_{-0.2}$ keV \\
& E$_F$ = $113^{+54}_{-30}$ keV & E$_F$ = $12.0^{+1.5}_{-1.5}$ keV & E$_F$ = $19.1^{+4.0}_{-3.0}$ keV \\
&  &  &  \\

&  &  &  \\
E$_C$ = 14 keV & E$_C$ = $9.3^{+0.4}_{-0.5}$ keV & E$_C$ = $13.4^{+0.5}_{-0.5}$ keV & E$_C$ = $14.7^{+2.0}_{-1.5}$ keV \\
& E$_F$ = $155^{+80}_{-39}$ keV & E$_F$ = $15.3^{+7.1}_{-4.7}$ keV & E$_F$ = $4.3^{+11.2}_{-4.3}$ keV \\
&  &  &  \\

&  &  &  \\
E$_C$ = 16 keV & E$_C$ = $8.1^{+0.8}_{-1.0}$ keV & E$_C$ = $14.3^{+0.9}_{-1.2}$ keV & E$_C$ = $8.0^{+0.6}_{-1.0}$ keV \\
& E$_F$ = $248^{+170}_{-93}$ keV & E$_F$ = $23^{+39}_{-12}$ keV & E$_F$ = $261^{+190}_{-79}$ keV \\
&  &  &  \\
\noalign{\smallskip}\hline
\end{tabular}
\end{table}

In Tab.~\ref{tab:hmxbtab2} are reported the high energy cut-off parameters obtained fitting the simulated spectra. The use of the CryoAC as hard X-ray detector allows to improve the characterization of the cut-off and the folding energies of the sources for E$_C$ $>$ 10 keV, whereas the TES array is unable to properly constrain them. Furthermore, note that also in this case the energy resolution of the CryoAC plays a fundamental role in the parameter characterization, significantly influencing the instrument performances.  

Finally, in order to evaluate the relevance of this case study in the context of the ATHENA mission, we have analyzed the last ATHENA Mock Observing Plan searching for bright HMXBs.  Within the planned X-IFU observations we have found: 
\begin{itemize}
\item 25 HMXB with average source intensity F$_{AVG} >$ 100 mCrab and planned exposure time t = 20 ks
\item 10 HMXB with 10 mCrab $< $ F$_{AVG} <$ 100 mCrab and planned exposure time in the range t = 20 - 200 ks
\end{itemize}
We can conclude that with a CryoAC energy resolution of 2 keV (FWHM) could be possible to characterize these source in the enlarged band up to 20 keV.

\section{Conclusions}
\label{sec:5}

We have performed a study aimed to understand if the CryoAC detector onboard ATHENA X-IFU can be optimized in order to extend the instrument scientific capability in the hard X-ray band.

We have found that in the baseline configuration the CryoAC could operate as hard X-ray detector in the narrow band 10-20 keV, with a limit flux (5$\sigma$, 100 ks, 1 pixel) of 1.6$\cdot$10$^{-12}$ erg/cm$^2$/s ($\sim$ 0.2 mCrab). The energy band is limited by the drop of the optics effective area at high energies, and not by the detector features.

Furthermore, we have found that an optimization of the CryoAC energy resolution could have a scientific return in the observation of bright sources with a spectral cut-off in this band. In this context we present the observation of HMXBs as case study, finding about 35 sources of this type in the ATHENA Mock Observing Plan. The required value that we have identified for the CryoAC energy resolution is $\Delta$E = 2 keV (FWHM) in the 10-20 keV band.

We remark that the intrinsic resolution of the CryoAC is estimated to be about one order of magnitude lower than this requirement: $\Delta$E $\sim$ 0.3 keV (FWHM) @ 15 keV. This value is given by the combination of two components: the electron-hole number fluctuation in the Silicon absorber and the usual thermal contribution related to the system thermal capacity. At present, the main uncertainty in this estimate comes from the evaluation of the absorber heat capacity, which strongly depends on the concentration of free electrons in the sample, i.e. on Silicon doping and resistivity \cite{silicon}. In this context, an activity to deeply characterize the Silicon wafer to be used for the detector production has been inserted in the CryoAC development schedule.  However, given the large margin, the proposed requirement $\Delta$E = 2keV (FWHM) seems full achievable without significant change in the baseline design.

\begin{acknowledgements}
The research leading to these results has received funding from the European Union's Horizon 2020 Programme under the AHEAD project (grant agreement n. 654215).
\end{acknowledgements}

\end{document}